\documentclass[sigconf]{acmart}

\renewcommand\footnotetextcopyrightpermission[1]{}
\setcopyright{none}

\acmConference[40th Brazilian Symposium on Software Engineering]{40th Brazilian Symposium on Software Engineering}{2026}{São Paulo, SP, Brazil}

\AtBeginDocument{
    
}

\usepackage{graphicx}
\usepackage{booktabs}
\usepackage{float}

\begin{document}

\title{Comparing the Quality of Code Generated by Vibe Coding Tools}

\author{Gustavo da Mota}
\affiliation{
  \institution{Universidade Federal de Pernambuco (UFPE)}
  \city{Recife}
  \country{Brazil}
}
\email{gpsm@cin.ufpe.br}

\author{Kiev Gama}
\affiliation{
  \institution{Universidade Federal de Pernambuco (UFPE)}
  \city{Recife}
  \country{Brazil}
}
\email{kiev@cin.ufpe.br}

\begin{abstract}
The use of AI agents for automatic code generation has become increasingly common in software development. However, concerns remain about the quality of the generated code, including aspects of maintainability, readability, and long-term evolution. This study compares the structural quality of code produced by three widely adopted vibe coding tools --- Lovable, v0, and Replit --- starting from a single generation prompt. We generate three independent projects per tool, totalling nine web applications, and submit them to static analysis with SonarQube. We collect metrics such as the number of issues, severity distribution, estimated remediation effort, cyclomatic and cognitive complexity, and code duplication. Preliminary results show that the tools exhibit distinct qualitative profiles: Lovable concentrates issues of lower severity but presents a substantially higher density of code smells per KLOC, while v0 and Replit produce more code with more aggressive severity profiles. These findings suggest that choosing between vibe coding tools involves structural trade-offs that go beyond perceived productivity.
\end{abstract}


\frenchspacing
\maketitle

\section{Introduction}

With the growing adoption of AI-based code generation tools in software development --- whose usage rose from 76\% in 2024 to 84\% in 2025 according to the Stack Overflow Developer Survey \cite{stackoverflow2025} --- it becomes possible to develop entire applications from a small set of natural-language instructions. This new possibility brings not only potential gains in productivity, reported by 69\% of AI-agent users \cite{stackoverflow2025}, but also concerns about the quality, readability, organization, and maintainability of the generated code~\cite{fawzy2025vibecoding, siddiq2022codesmells}. Empirical studies reinforce these concerns: Pearce et al.~\cite{pearce2022asleep} found that around 40\% of the programs completed by GitHub Copilot in security-relevant scenarios contained vulnerabilities, while evaluations spanning multiple tools report wide variation in the correctness, security, and maintainability of the generated code~\cite{yetistiren2023evaluating}.

Already consolidated in software engineering for analyzing code written by human developers, static analysis can serve as an important instrument to assess code quality. As reported by Trautsch et al.~\cite{trautsch2023asat}, the accumulation of issues identified by static analysis tools tends to be associated with a higher number of real problems during software execution.

In this context, this study proposes to compare AI-based application generation tools through the static analysis of the code they produce. We seek to identify differences between the evaluated tools, as well as recurring patterns of structural problems present in the produced code.

\section{Background}
\subsection{Vibe Coding and AI-Assisted Code Generation}

The notion of vibe coding has been used to describe an emerging way of building software, in which the developer interacts in natural language with AI tools that take over most of the structural decisions of the system. The term was popularized by Andrej Karpathy in 2025, when he described a practice in which the developer \textit{``gives in''} to the flow of interaction with generative models and steers the system more by intent than by the literal writing of code~\cite{karpathy2025vibe}. Recent surveys highlight the importance of evaluating other aspects of the code (e.g., security, maintenance, readability, and adherence to good practices), and point to a dichotomy between speed and quality, with users reporting a considerable trade-off between speed of \textit{``functionality''} and structural and maintenance inconsistencies~\cite{jiang2024survey, fawzy2025vibecoding}. Unlike traditional programming assistants, whose main focus is to help writing or editing code, platforms such as Lovable, v0, and Replit AI propose development flows with a high level of abstraction and automated application generation~\cite{fawzy2025vibecoding}. This widens the research gap in software engineering: instead of evaluating isolated functions or algorithmic solutions, it became necessary to examine the structural quality of entire AI-generated applications.

\subsection{Code Quality, Code Smells, and Tech Debt}

The notion of code quality involves multiple dimensions, and different analysis tools vary in the relevance they assign to each evaluation aspect. Models such as ISO/IEC 25010 separate these aspects into categories such as maintainability, reliability, security, performance efficiency, and portability.

Fowler et al.~\cite{fowler1999refactoring} introduced the concept of code smells, which comprises surface code patterns that may indicate deeper problems of maintenance, comprehensibility, and defects in the system. The relevance of code smells stems from the fact that they do not necessarily cause immediate failures, but can increase complexity, hinder maintenance, and contribute to the accumulation of technical debt. Tech debt can in turn be understood as the future cost --- in time and effort --- associated with decisions that facilitate short-term delivery but produce instability and difficulty of future maintenance. Static analysis tools such as SonarQube aim to detect part of these concepts through automated metrics and rules. SonarQube classifies problems into categories such as bugs, vulnerabilities, and code smells, defining a code smell as a maintainability problem that makes the code confusing and hard to maintain. The tool also computes metrics such as technical debt and technical debt ratio, associating an estimated remediation effort with the issues found. As shown by Trautsch et al.~\cite{trautsch2023asat}, the accumulation of issues reported by static analysis usually indicates a higher number of real problems in the system.

Thus, using static analysis tools to detect code smells in applications generated by vibe coding tools, while not replacing functional testing and manual review, represents a first step to investigate the structural quality of the code.

\section{Methodology}

Rather than declaring one tool superior to the others, the goal of this study is to obtain initial evidence about the structural quality of code generated by vibe coding tools from the same requirements specification. Adopting an exploratory comparative design, suitable for the Innovative Ideas and Emerging Results track, we aim to answer the following research question:

\textbf{RQ1:} \textit{How do vibe coding tools differ in the number and severity of code smells detected in web applications generated from the same set of requirements?}

The unit of analysis is a project generated by a web application generation platform, fulfilling the functional requirements stated in a prompt that is identical for all interactions with the models.

\subsection{Prompt Design}

The prompt was designed to elicit applications with a minimum level of complexity, while not providing the model with examples of previous implementations to draw upon. This approach, known as zero-shot~\cite{brown2020fewshot}, is characterized by the absence of code or prior implementations in the prompt, so that the model produces a response based exclusively on the supplied instructions.

The requirements were based on an example application called \textit{``Nature Park Wildlife App''}~\footnote{This activity was originally developed for Michel Chaudron's classes at Eindhoven University of Technology and also used in a hands-on session of the 3rd Designing Workshop at ICSE 2026}, in which users can register wildlife sightings, report safety issues, and subscribe to notifications about specific animals, supported by a geolocated map. The full prompt is included in the research artifacts.

\subsection{Tools Under Evaluation}

The selected tools were Lovable, v0, and Replit. They are three widely used platforms, completely web-based, that aim to generate entire applications from a few natural-language interactions and that offer free tiers with token quotas sufficient to support the experiments.

As these are continuously updated SaaS platforms, the tools do not expose public version numbers; we therefore report the generation window of the projects --- between May 6 and May 21, 2026 --- together with the build identifiers observable in the artifacts. On Lovable, the projects were generated from the \texttt{tanstack\_start\_ts} template (dated 2026-05-06 and 2026-05-12) by the \texttt{gpt-engineer-app} agent; on v0, by the \textit{v0.app} platform, producing Next.js 16.2.4 applications; and on Replit, by the \textit{Replit Agent} (Nix channel \texttt{stable-25\_05}, Node.js 24).

\subsection{Generation Procedure}

Each tool was exposed to the same prompt, with no additional instructions or manual corrections. The outcome of each iteration was documented and the projects were stored in separate repositories. The process was repeated three times per tool, in order to minimize the influence of atypical outputs on the final results, which are possible due to the non-deterministic nature of the models.

This procedure yielded nine distinct projects:

\begin{itemize}
  \item \textbf{Lovable:}\\
\texttt{wildlife-whisperer-app},\\
 \texttt{wild-park-patrol},\\
    \texttt{nature-watch-live}.
  \item \textbf{v0:}\\
 \texttt{v0-aplicativo-de-rastreamento},\\
 \texttt{v0-wildlife-tracking-app},\\
    \texttt{v0-wildlife-track-app}.
  \item \textbf{Replit:}\\
    \texttt{Wildlife-Tracker-replit-1},\\
    \texttt{Wildlife-Tracker-replit-2},\\
    \texttt{Nature-Watch-System-replit-3}.
\end{itemize}

\subsection{Static Analysis Procedure}

The evaluation was independent for each project and used SonarQube as the static analysis tool. SonarQube is among the most widely adopted static analysis tools in both industry and software engineering research~\cite{avgeriou2021td, lenarduzzi2023comparison}, which favors the comparability and reproducibility of our results. SonarQube produces a list of issues found in the project. Each issue contains: violated rule, severity, source component, line, message, type, tags, and remediation effort. In addition, project-level information is produced, such as duplicated-line density, cyclomatic complexity, cognitive complexity, and non-commented lines of code~\cite{sonarqube_docs}.

\subsubsection{Rules, Tags, and Issue Types}

The rule is more literal regarding which basic rule was violated for that issue to be raised, while tags are broader signifiers that give meaning to the problem itself and can be more than one per issue. For instance, the use of a deprecated API will trigger rule S1874, and the tags of that issue will include the category \textit{obsolete}. Issue types fall into three main categories: code smells, bugs, and vulnerabilities.

\subsubsection{Remediation Effort}

The effort metric represents an estimated time to remediate the issue, providing a possible measure of technical debt. From the available data, we can not only compare the tools numerically, but also compare qualitatively the type of problem each tool is more prone to generate, as well as their severity and remediation effort.

\subsection{Comparison Strategy}

For comparison, we primarily use the number of issues per project, the associated effort, and the proportional distribution of their types and tags.

\subsubsection{Quantitative Comparison}

The quantitative comparison has two axes: raw and proportional. Each tool produced a different volume of code, which makes a proportional analysis necessary to avoid misleading conclusions from raw numbers. On the other hand, the expected functionality of the projects is the same, which makes the non-proportional analysis also relevant, as it is, in a sense, proportional to the delivered functionality.

\subsubsection{Qualitative Comparison}

The qualitative comparison examines, within each project, the predominant
patterns of each tool. To this end, we primarily rely on the severity
distribution of issues and, as a complementary indicator, on their tags,
chosen for their descriptive nature. The tags generated by the issues of
the nine projects amount to thirty-nine distinct categories, which were
grouped into five clusters to improve data visualization: Semantic,
Technical Quality, Security, Technological Context, and Language /
Ecosystem. The complete mapping of tags into these clusters is presented in
Table~\ref{tab:tags-en}.

\begin{table}[ht]
  \caption{Mapping of tags into clusters.}
  \label{tab:tags-en}
  \footnotesize
  \begin{tabular}{lp{5.5cm}}
    \toprule
    Cluster & Tags \\
    \midrule
    Semantic
      & readability, brain-overload, confusing, redundant,
        suspicious, pitfall, consistency, unused,
        error-handling, bad-practice \\
    \midrule
    Technical Quality
      & performance, accessibility, optimization \\
    \midrule
    Security
      & cwe \\
    \midrule
    Technological Context
      & react, jsx, html, javascript, regex, async,
        import, esm-only, eslint, wcag, nodejs, linting \\
    \midrule
    Language / Ecosystem
      & es2015, es2020, es2021, es2022,
        internationalization, portability, formatting,
        convention, type-dependent, obsolete,
        nullish-coalescing, modernize, unicode \\
    \bottomrule
  \end{tabular}
\end{table}

\section{Data Presentation}

This section presents the data extracted from the static analysis of the nine generated projects. For each evaluated tool, we detail the per-project information (and the aggregated total), followed by the profiles of issue types, severity, tags, and rules. Section~\ref{sec:tool-comparison-en} then consolidates the data into a direct comparison between the three tools, in raw values and normalized per KLOC.

Before the per-tool breakdown, Table~\ref{tab:rules-en} describes the SonarQube rules that appear in the rule-frequency charts of the three tools (Figures~\ref{fig:rules-lovable-en}, \ref{fig:rules-v0-en}, and~\ref{fig:rules-replit-en}), so that the subsequent analysis can refer to them directly. The descriptions follow the official SonarQube rule documentation~\cite{sonarqube_rules}.

\begin{table}[ht]
  \caption{SonarQube most recurring rules appearing in the per-tool rule-frequency charts (\texttt{typescript:} prefix omitted).}
  \label{tab:rules-en}
  \footnotesize
  \setlength{\tabcolsep}{4pt}
  \begin{tabular}{@{}lp{6.0cm}@{}}
    \toprule
    Rule & Description\\
    \midrule
    \texttt{S1874} & Deprecated APIs, classes, or functions should not be used.\\
    \texttt{S3358} & Ternary operators should not be nested.\\
    \texttt{S4325} & Redundant casts and non-null assertions should be removed.\\
    \texttt{S6481} & React Context Provider values should have stable identities.\\
    \texttt{S6759} & React component props should be read-only.\\
    \texttt{S6819} & Prefer a semantic HTML tag over an ARIA \textit{role}.\\
    \texttt{S7735} & Negated conditions should be avoided when an \textit{else} clause is present.\\
    \texttt{S7748} & Number literals should not have unnecessary decimal points or trailing zeros.\\
    \texttt{S7764} & Use \texttt{globalThis} instead of \texttt{window}, \texttt{self}, or \texttt{global}.\\
    \texttt{S7773} & \texttt{Number} static methods and properties should be preferred over their global equivalents.\\
    \texttt{S7781} & Use \texttt{replaceAll()} instead of \texttt{replace()} with a global \textit{regex}.\\
    \bottomrule
  \end{tabular}
\end{table}

\subsection{Lovable}

Table~\ref{tab:lovable-en} summarizes the raw metrics for the three projects generated by Lovable. The tool produced the lowest total volume of code among the three evaluated tools and introduced only a single bug according to the SonarQube classification, but it concentrated the highest density of code smells per line of code.

\begin{table}[ht]
  \caption{Per-project metrics --- Lovable.}
  \label{tab:lovable-en}
  \footnotesize
  \setlength{\tabcolsep}{4pt}
  \begin{tabular}{lrrrr}
    \toprule
    Project & NCLOC & Issues & Eff.(min) & Cog.Cplx\\
    \midrule
    wildlife-whisperer-app &  5,781 &  59 &   215 & 225\\
    wild-park-patrol       &  5,697 & 167 & 1,782 & 237\\
    nature-watch-live      &  5,603 & 164 & 1,774 & 219\\
    \midrule
    \textbf{Total}     & \textbf{17,081} & \textbf{390} & \textbf{3,771} & \textbf{681}\\
    \bottomrule
  \end{tabular}
\end{table}

As shown in the table, the \texttt{wildlife-whisperer-app} project yielded considerably different results from the other two projects, at least in the issues and effort metrics. This divergence is plausible and consistent with the discussion of model non-determinism, and reinforces the importance of multiple iterations per tool to minimize the impact of anomalous outputs.

\begin{figure}[ht]
  \centering
  \includegraphics[width=\linewidth]{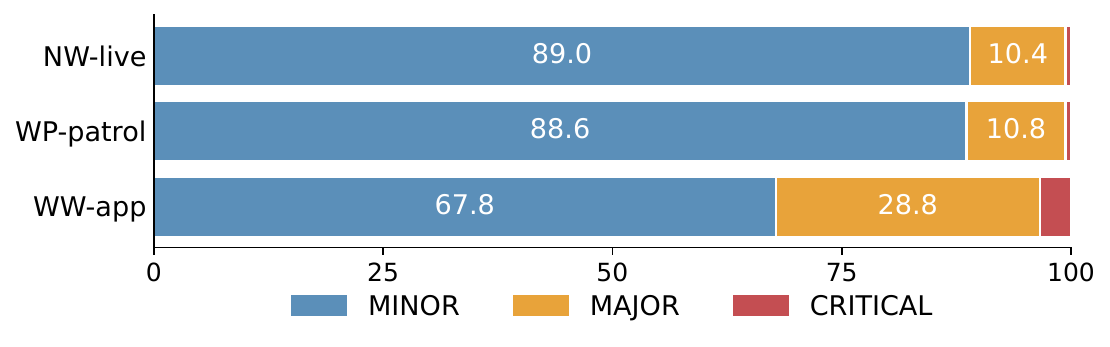}
  \caption{Severity distribution per project --- Lovable (\%).}
  \label{fig:sev-proj-lovable-en}
\end{figure}

A characteristic of Lovable is that the vast majority of its issues were classified as \textit{Code Smell} (389 out of 390), in contrast to the other tools, which presented more varied type profiles. In addition, the severity of the detected problems was mostly MINOR, as can be seen in Figure~\ref{fig:sev-proj-lovable-en}, with \texttt{wildlife-whisperer-app} once again showing a large gap with respect to the other projects of the tool.

\begin{figure}[ht]
  \centering
  \includegraphics[width=\linewidth]{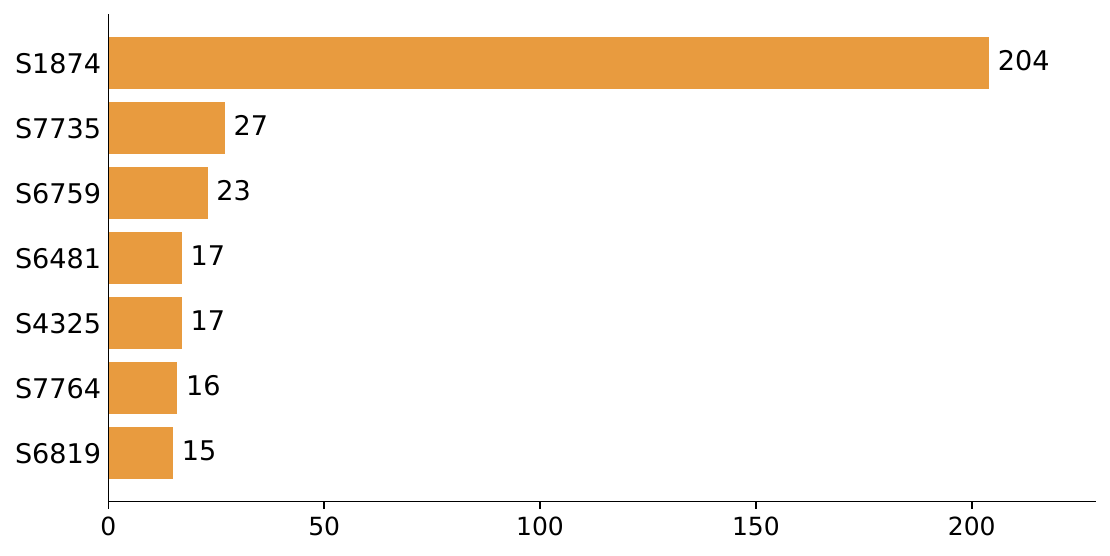}
  \caption{Most frequent rules in Lovable's issues (top~7, \texttt{typescript:} prefix omitted).}
  \label{fig:rules-lovable-en}
\end{figure}

The most striking aspect of Lovable's data is the predominance of issues raised by rule \texttt{S1874}, which targets obsolete APIs and deprecated classes and functions (Table~\ref{tab:rules-en}). As Figure~\ref{fig:rules-lovable-en} shows, this signature will also be visible in the cross-tool comparison.

\subsection{v0}

Table~\ref{tab:v0-en} summarizes the raw metrics for the three projects generated by v0. The tool produced an intermediate code volume and the lowest total remediation effort, with only a few issues classified as bugs.

\begin{figure}[ht]
  \centering
  \includegraphics[width=\linewidth]{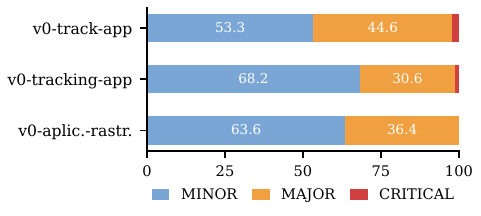}
  \caption{Severity distribution per project --- v0 (\%).}
  \label{fig:sev-proj-v0-en}
\end{figure}

\begin{table}[ht]
  \caption{Per-project metrics --- v0.}
  \label{tab:v0-en}
  \footnotesize
  \setlength{\tabcolsep}{3pt}
  \begin{tabular}{lrrrr}
    \toprule
    Project & NCLOC & Issues & Eff.(min) & Cog.Cplx\\
    \midrule
    v0-aplicativo-de-rastreamento &  8,790 &  77 & 284 & 277\\
    v0-wildlife-tracking-app      &  9,278 &  85 & 283 & 305\\
    v0-wildlife-track-app         &  9,515 &  92 & 374 & 354\\
    \midrule
    \textbf{Total}     & \textbf{27,583} & \textbf{254} & \textbf{941} & \textbf{936}\\
    \bottomrule
  \end{tabular}
\end{table}

v0 positioned itself as intermediate on several metrics, which is especially visible in the lines-of-code count: all three projects stayed close to nine thousand lines, with low standard deviation across projects. Regarding severity (Figure~\ref{fig:sev-proj-v0-en}), it also remained intermediate in the overall proportion of severity levels.

\begin{figure}[ht]
  \centering
  \includegraphics[width=\linewidth]{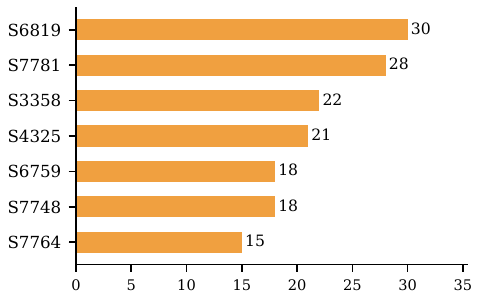}
  \caption{Most frequent rules in v0's issues (top~7, \texttt{typescript:} prefix omitted).}
  \label{fig:rules-v0-en}
\end{figure}

Regarding the rules most violated by v0's code, the most salient feature is the more distributed profile: no single problem dominates as in Lovable. As for the nature of the issues, they relate more to framework-specific best practices, in addition to rules \texttt{S7735} and \texttt{S7748}, which capture readability and numeric-literal formatting problems (Figure~\ref{fig:rules-v0-en}; Table~\ref{tab:rules-en}).

\subsection{Replit}

Table~\ref{tab:replit-en} presents the raw metrics for the three projects generated by Replit. The tool produced the largest total volume of code, with substantial line duplication and the broadest coverage of distinct rules (35).

\begin{table}[ht]
  \caption{Per-project metrics --- Replit.}
  \label{tab:replit-en}
  \footnotesize
  \setlength{\tabcolsep}{3pt}
  \begin{tabular}{lrrrr}
    \toprule
    Project & NCLOC & Issues & Eff.(min) & Cog.Cplx\\
    \midrule
    Wildlife-Tracker-replit-1    & 16,827 & 177 & 632 & 655\\
    Wildlife-Tracker-replit-2    & 15,736 & 140 & 520 & 558\\
    Nature-Watch-System-replit-3 & 15,848 & 146 & 536 & 602\\
    \midrule
    \textbf{Total} & \textbf{48,411} & \textbf{463} & \textbf{1,688} & \textbf{1,815}\\
    \bottomrule
  \end{tabular}
\end{table}

\begin{figure}[ht]
  \centering
  \includegraphics[width=\linewidth]{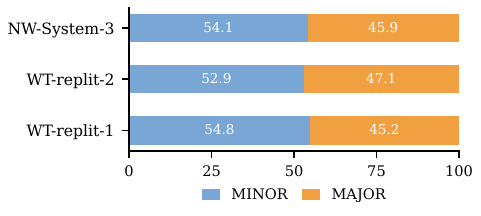}
  \caption{Severity distribution per project --- Replit (\%).}
  \label{fig:sev-proj-replit-en}
\end{figure}

At first glance, the high number of lines of code produced by Replit stands out. This is also reflected in the number of duplicated lines, with an average of about 60\% --- a remarkably high figure considering that the next tool with most duplication, v0, never exceeded 7\%. Another negative metric was the severity of the generated problems: Replit produced, among the three tools, the highest absolute and relative share of severe issues (Figure~\ref{fig:sev-proj-replit-en}).

\begin{figure}[ht]
  \centering
  \includegraphics[width=\linewidth]{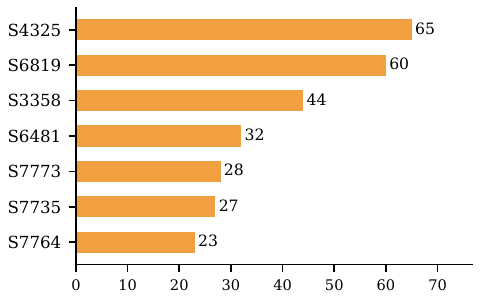}
  \caption{Most frequent rules in Replit's issues (top~7, \texttt{typescript:} prefix omitted).}
  \label{fig:rules-replit-en}
\end{figure}

Regarding the rules violated in Replit's projects, two stand out (Figure~\ref{fig:rules-replit-en}): \texttt{S4325}, related to redundant type assertions, and \texttt{S6819}, an accessibility rule that recommends semantic HTML elements instead of ARIA \textit{role} attributes (Table~\ref{tab:rules-en}).

\subsection{Tool Comparison}
\label{sec:tool-comparison-en}

For the cross-tool comparison, it is important to present the data both in absolute form and in relative form --- in this case, per thousand lines of code. Raw figures may seem to favor tools that produce less code (and thus less surface for errors), but they remain essential: there is no merit in fulfilling the same functional requirements with more lines of code, especially when there is massive duplication, as in the case of Replit.

Table~\ref{tab:comparativo-bruto-en} consolidates, per tool, the raw totals and the same values normalized per KLOC (thousand non-commented lines of code). The effort metric is expressed in estimated remediation minutes.

\begin{table*}[t]
  \caption{Comparison across tools: raw totals and KLOC-normalized values.}
  \label{tab:comparativo-bruto-en}
  \footnotesize
  \setlength{\tabcolsep}{6pt}
  \begin{tabular}{lrrrrrrr}
    \toprule
    Tool & NCLOC & Issues & CS & Bugs & Hot. & Eff.(min) & D.Rules\\
    \midrule
    Lovable & 17,081 & 390 & 389 & 1 &  6 & 3,771 & 24\\
    v0      & 27,583 & 254 & 249 & 5 & 10 &   941 & 24\\
    Replit  & 48,411 & 463 & 455 & 8 & 30 & 1,688 & 35\\
    \midrule
    \multicolumn{8}{l}{\textit{Normalized per KLOC}}\\
    \midrule
    Lovable & --- & 22.83 & 22.77 & 0.06 & 0.35 & 220.77 & ---\\
    v0      & --- &  9.21 &  9.03 & 0.18 & 0.36 &  34.12 & ---\\
    Replit  & --- &  9.56 &  9.40 & 0.17 & 0.62 &  34.87 & ---\\
    \bottomrule
  \end{tabular}
\end{table*}

\subsubsection{Remediation Effort}

\begin{figure}[ht]
  \centering
  \includegraphics[width=\linewidth]{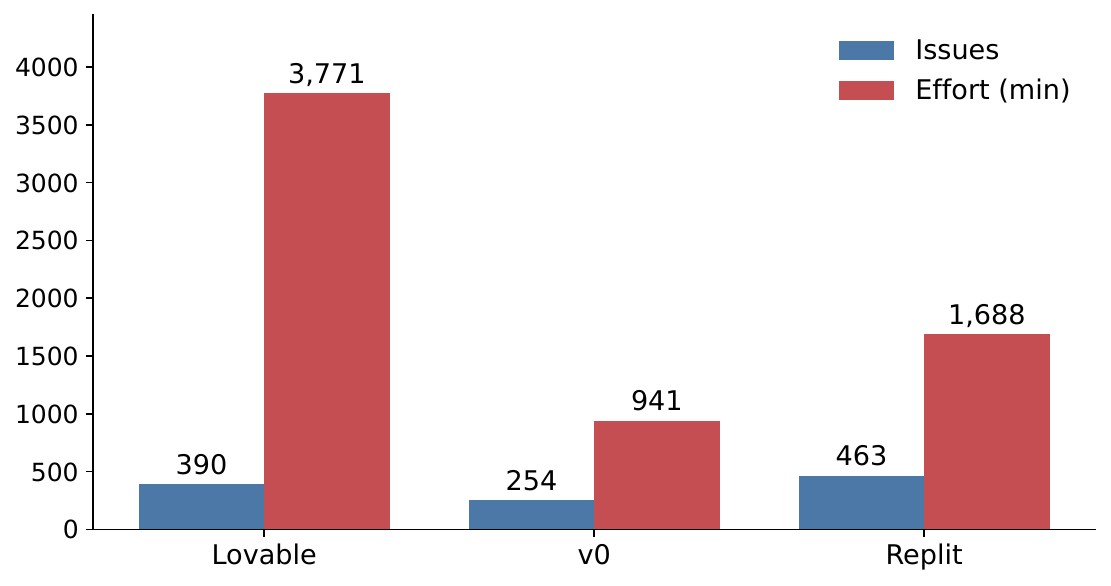}
  \caption{Absolute comparison: total number of issues and total remediation effort (in minutes) per tool.}
  \label{fig:comp-abs-en}
\end{figure}

Figure~\ref{fig:comp-abs-en} contrasts the total number of issues and the total remediation effort per tool, in absolute terms. Although Lovable has fewer issues than Replit, it has the largest remediation-time requirement, showing that the lower severity of Lovable's issues does not translate into low correction effort.

\subsubsection{Qualitative Profile of Issues}

\begin{figure}[ht]
  \centering
  \includegraphics[width=\linewidth]{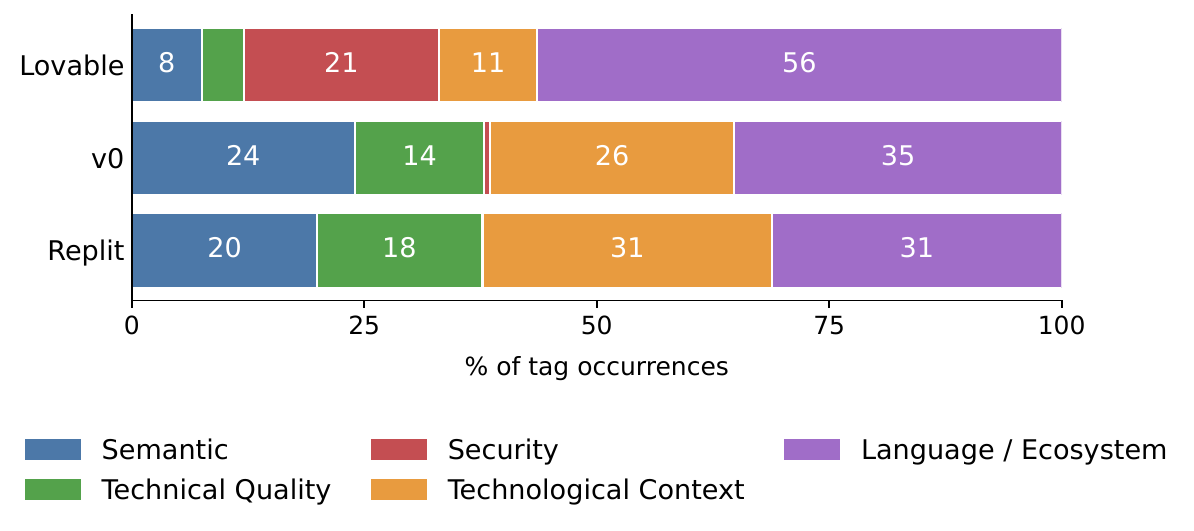}
  \caption{Tag profile by \textit{cluster} (\% per tool).}
  \label{fig:comp-macro-en}
\end{figure}

Figure~\ref{fig:comp-macro-en} shows the tag profile aggregated by \textit{cluster}, in proportion (\%), allowing a qualitative comparison of which structural dimensions each tool concentrates its problems in. Lovable stands out for the heaviest concentration on the security cluster and, above all, on the language / ecosystem cluster --- the latter strongly driven by rule \texttt{S1874}. Replit and v0 show qualitative profiles similar to each other.

\begin{figure}[ht]
  \centering
  \includegraphics[width=\linewidth]{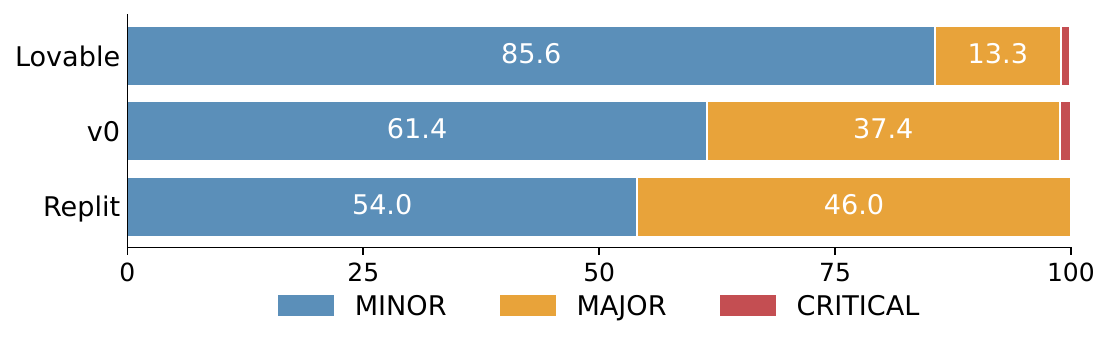}
  \caption{Severity distribution per tool (percentages).}
  \label{fig:comp-sev-en}
\end{figure}

As for the severity distribution among the three tools in percentages (Figure~\ref{fig:comp-sev-en}), the shift from Lovable's profile (predominantly MINOR) to Replit's profile (nearly balanced between MINOR and MAJOR) is evident, with Replit being the only tool not producing any CRITICAL-classified problem. Lovable and v0 produce small shares of CRITICAL issues (1.0\% and 1.2\% respectively).

\section{Discussion}

The findings of this study highlight a trade-off that is practically invisible
in the adoption of vibe coding tools: reported productivity gains do not
necessarily reflect structural quality. The Lovable case illustrates this
paradox --- the tool presented an intermediate issue count in static analysis,
yet the highest remediation effort among the evaluated tools, suggesting that
surface-level metrics may induce a false perception of quality.

The predominance of rule \texttt{S1874} in the projects generated by Lovable
--- deprecated APIs accounting for 204 occurrences --- finds an explanation in
the literature on LLMs for code generation. Models trained on large historical
corpora tend to over-represent API patterns that were prevalent at the time of
training, even when those patterns have since been formally discontinued. This
phenomenon, documented by Lin et~al.~\cite{lin2025deprecated} and by subsequent
studies on knowledge conflicts in LLMs, suggests that vibe coding tools may
systematically generate code oriented toward the past of the ecosystem --- not
toward its current state.

This risk takes on an additional dimension when the target audience of these
platforms is considered. Tools such as Lovable are explicitly positioned for
non-developer users, lowering the barrier to creating complete web applications.
However, this same accessibility eliminates the technical review layer that, in
traditional development contexts, would serve as a filter for structural and
security issues. Documented incidents --- such as \texttt{CVE-2025-48757},
which exposed user data in more than 170 applications generated by the
platform~\cite{superblocks2025lovable,cve2025lovable} --- illustrate the
concrete consequences of this gap. The present study contributes precisely by
making this trade-off measurable: not as a criticism of the tools themselves,
but as evidence that their adoption requires awareness of the structural debt
profiles that each tends to produce.

\section{Limitations}
\label{sec:limitations}

The main methodological challenge of this study refers to the capacity of static analysis to sufficiently quantify the code quality of a software project. Recent literature positions SonarQube as a relevant auxiliary indicator, but not as an instrument capable of definitively assessing code quality. Marcilio et al.~\cite{marcilio2019sonarqube}, in a study spanning 421,976 issues from 246 projects across four distinct SonarQube instances, observed that on average only 13\% of the issues reported by the tool are effectively fixed --- and conjecture that only a subset of its checkers actually reveal real design and coding flaws, which may artificially inflate the measured technical-debt index.

A second limitation concerns the sampling scope of the study. Due to token-credit restrictions in the free versions of the tools and the project's time horizon, only three projects were generated per tool --- a volume still susceptible to anomalous outputs. The collected data illustrate this risk concretely: in the set of projects generated by Lovable, the standard deviation of the code-smells metric was 62.1, considerably higher than that observed for v0 (SD = 7.0) and Replit (SD = 20.3). This dispersion is explained, to a large extent, by the project \texttt{wildlife-whisperer-app}, which recorded only 58 code smells --- a substantial divergence from the other projects of the same tool (164 and 167, respectively) --- and which, on its own, disproportionately distorts the tool's mean. With larger samples, such fluctuations would tend to be diluted, making the comparisons between tools more robust and less sensitive to point variations of a single execution.

Finally, a third limitation concerns the opacity of the platforms under evaluation: there is no public transparency about how each platform allocates computational resources among its users in the free tiers. It is plausible to speculate --- and this is speculation, not a claim --- that, at peak times, lower-capacity models may be invoked in place of the platform's default model, affecting the quality of the generated code. This factor reinforces the need for broader replications across varied contexts before drawing definitive conclusions.

\section{Conclusion and Future Work}

This study aimed to provide initial evidence that can guide the choice between vibe coding tools from the perspective of the structural quality of the generated code. Taken together, the data suggest that each of the three evaluated tools presents its own risk profile, with distinct implications for adopting them in a real development workflow. Among the three, v0 emerged as the most balanced alternative. It produced the lowest total remediation effort (941 minutes, against 3,771 for Lovable and 1,688 for Replit), and intermediate severity proportions, with neither the peak of MAJOR issues observed in Replit nor the strong concentration on deprecated APIs observed in Lovable.

Lovable exposes a relevant trade-off: although the severity of its issues is largely MINOR ($\sim$86\%), the tool concentrated the highest estimated remediation effort ($\sim$4.0$\times$ v0's), with more than half of that effort tied to a single rule (\texttt{S1874}, on deprecated APIs and functions; Table~\ref{tab:rules-en}). This specific concentration, combined with the weight of the \textit{cwe} tag on the security cluster, suggests that Lovable may produce code that looks clean but relies on APIs whose reliability --- including in security terms --- has already been questioned by the ecosystem itself. It should be noted, as discussed in Section~\ref{sec:limitations}, that part of Lovable's average is pulled by the \texttt{wildlife-whisperer-app} project, whose divergent behavior relative to the other Lovable projects calls for caution when interpreting these averages in isolation.

Replit, in turn, presents two characteristics that deserve special attention. The first is the massive line duplication, with an average of approximately 60\% across all three projects --- a figure substantially above v0 ($\leq$7\%) and Lovable (0\%), indicating strong structural verbosity. The second is the highest absolute and proportional share of MAJOR-severity problems (213 occurrences, 46\% of the total), above the other two tools. Combined, these two factors suggest that the productivity gain from using Replit may come with a more significant structural debt over the medium term.

As future work, we intend to: (i) significantly increase the number of iterations per tool, mitigating the effect of anomalous outputs such as the one observed in \texttt{wildlife-whisperer-app}; (ii) submit the \textit{premium} versions of these platforms to the same procedure, under the hypothesis that differences in model tier may materially affect the results; and (iii) incorporate other static analysis tools besides SonarQube, in order to reduce the bias of single-tool coverage and make the comparative reading more robust.

\section*{Artifact Availability}
The research artifacts --- complete prompt used in every generation, source code of the nine generated projects and full SonarQube reports --- are available at \url{https://anonymous.4open.science/r/comparing-vibe-coding-tools-artifacts-45EE/} (anonymized for review).

\section*{Acknowledgements}
Claude was used to generate all the figures and tables in this paper, based on explicit specifications and guidance from the authors. Additionally, Claude assisted in editing and improving the overall quality of the text, contributing to clarity, grammar, and structure.
For partially supporting this work, we would like to thank INES.IA (National Institute of Science and Technology for Software Engineering Based on and for Artificial Intelligence) www.ines.org.br, CNPq grant 408817/2024-0.

\bibliographystyle{ACM-Reference-Format}
\bibliography{sample-base}

@misc{stackoverflow2025,
  author       = {{Stack Overflow}},
  title        = {Stack Overflow Developer Survey 2025},
  year         = {2025},
  howpublished = {\url{https://survey.stackoverflow.co/2025/}},
  note         = {Accessed 2026-05}
}

@book{fowler1999refactoring,
  author    = {Martin Fowler and
               Kent Beck and
               John Brant and
               William Opdyke and
               Don Roberts},
  title     = {Refactoring: Improving the Design of Existing Code},
  edition   = {1},
  publisher = {Addison-Wesley Professional},
  address   = {Boston, MA},
  year      = {1999},
  isbn      = {0-201-48567-2}
}

@article{trautsch2023asat,
  author    = {Alexander Trautsch and
               Steffen Herbold and
               Jens Grabowski},
  title     = {Are Automated Static Analysis Tools Worth It? {A}n Investigation
               into Relative Warning Density and External Software Quality on the
               Example of {Apache} Open Source Projects},
  journal   = {Empirical Software Engineering},
  volume    = {28},
  number    = {3},
  pages     = {66},
  year      = {2023},
  doi       = {10.1007/s10664-023-10301-2},
  url       = {https://doi.org/10.1007/s10664-023-10301-2},
  publisher = {Springer Nature}
}

@misc{fawzy2025vibecoding,
  author        = {Ahmed Fawzy and Amjed Tahir and Kelly Blincoe},
  title         = {Vibe Coding in Practice: Motivations, Challenges, and a Future
                   Outlook -- A Grey Literature Review},
  year          = {2025},
  eprint        = {2510.00328},
  archivePrefix = {arXiv},
  primaryClass  = {cs.SE},
  url           = {https://arxiv.org/abs/2510.00328},
  note          = {arXiv preprint}
}

@misc{karpathy2025vibe,
  author       = {Andrej Karpathy},
  title        = {There's a New Kind of Coding {I} Call ``{V}ibe Coding''},
  howpublished = {Post on X (formerly Twitter), \url{https://x.com/karpathy/status/1886192184808149383}},
  year         = {2025},
  month        = feb,
  note         = {Accessed 2026-05}
}

@misc{jiang2024survey,
  author        = {Juyong Jiang and
                   Fan Wang and
                   Jiasi Shen and
                   Sungju Kim and
                   Sunghun Kim},
  title         = {A Survey on Large Language Models for Code Generation},
  year          = {2024},
  eprint        = {2406.00515},
  archivePrefix = {arXiv},
  primaryClass  = {cs.SE},
  url           = {https://arxiv.org/abs/2406.00515},
  note          = {arXiv preprint}
}

@inproceedings{brown2020fewshot,
  author    = {Tom B. Brown and
               Benjamin Mann and
               Nick Ryder and
               Melanie Subbiah and
               Jared D. Kaplan and
               Prafulla Dhariwal and
               Arvind Neelakantan and
               Pranav Shyam and
               Girish Sastry and
               Amanda Askell and
               Sandhini Agarwal and
               Ariel Herbert-Voss and
               Gretchen Krueger and
               Tom Henighan and
               Rewon Child and
               Aditya Ramesh and
               Daniel M. Ziegler and
               Jeffrey Wu and
               Clemens Winter and
               Christopher Hesse and
               Mark Chen and
               Eric Sigler and
               Mateusz Litwin and
               Scott Gray and
               Benjamin Chess and
               Jack Clark and
               Christopher Berner and
               Sam McCandlish and
               Alec Radford and
               Ilya Sutskever and
               Dario Amodei},
  title     = {Language Models are Few-Shot Learners},
  booktitle = {Advances in Neural Information Processing Systems},
  volume    = {33},
  series    = {NeurIPS 2020},
  pages     = {1877--1901},
  year      = {2020},
  publisher = {Curran Associates, Inc.},
  note      = {\url{https://arxiv.org/abs/2005.14165}}
}

@inproceedings{pearce2022asleep,
  author    = {Hammond Pearce and Baleegh Ahmad and Benjamin Tan and
               Brendan Dolan-Gavitt and Ramesh Karri},
  title     = {Asleep at the Keyboard? {A}ssessing the Security of {GitHub} {Copilot}'s
               Code Contributions},
  booktitle = {2022 IEEE Symposium on Security and Privacy (SP)},
  pages     = {754--768},
  year      = {2022},
  doi       = {10.1109/SP46214.2022.9833571}
}

@inproceedings{siddiq2022codesmells,
  author    = {Mohammed Latif Siddiq and Shafayat H. Majumder and
               Maisha R. Mim and Sourov Jajodia and Joanna C. S. Santos},
  title     = {An Empirical Study of Code Smells in Transformer-based
               Code Generation Techniques},
  booktitle = {2022 IEEE 22nd International Working Conference on
               Source Code Analysis and Manipulation (SCAM)},
  pages     = {71--82},
  year      = {2022},
  doi       = {10.1109/SCAM55253.2022.00014}
}

@misc{yetistiren2023evaluating,
  author        = {Burak Yeti{\c{s}}tiren and I{\c{s}}{\i}k {\"O}zsoy and
                   Miray Ayerdem and Eray T{\"u}z{\"u}n},
  title         = {Evaluating the Code Quality of AI-Assisted Code Generation Tools:
                   An Empirical Study on {GitHub} {Copilot}, {Amazon} {CodeWhisperer},
                   and {ChatGPT}},
  year          = {2023},
  eprint        = {2304.10778},
  archivePrefix = {arXiv},
  primaryClass  = {cs.SE},
  url           = {https://arxiv.org/abs/2304.10778},
  note          = {arXiv preprint}
}

@article{avgeriou2021td,
  author  = {Paris C. Avgeriou and Davide Taibi and Apostolos Ampatzoglou and
             Francesca Arcelli Fontana and Terese Besker and
             Alexander Chatzigeorgiou and Valentina Lenarduzzi and
             Antonio Martini and Athanasia Moschou and Ilaria Pigazzini and
             Nyyti Saarim{\"a}ki and Darius Sas and
             Saulo Soares de Toledo and Angeliki-Agathi Tsintzira},
  title   = {An Overview and Comparison of Technical Debt Measurement Tools},
  journal = {IEEE Software},
  volume  = {38},
  number  = {3},
  pages   = {61--71},
  year    = {2021},
  doi     = {10.1109/MS.2020.3024958}
}

@article{lenarduzzi2023comparison,
  author  = {Valentina Lenarduzzi and Fabiano Pecorelli and
             Nyyti Saarim{\"a}ki and Savanna Lujan and Fabio Palomba},
  title   = {A Critical Comparison on Six Static Analysis Tools:
             Detection, Agreement, and Precision},
  journal = {Journal of Systems and Software},
  volume  = {198},
  pages   = {111575},
  year    = {2023},
  doi     = {10.1016/j.jss.2022.111575}
}

@misc{sonarqube_rules,
  author       = {{SonarSource}},
  title        = {{SonarQube}: {JavaScript}/{TypeScript} Static Analysis Rules},
  howpublished = {\url{https://rules.sonarsource.com/typescript/}},
  year         = {2025},
  note         = {Accessed 2026-05}
}

@inproceedings{marcilio2019sonarqube,
  author    = {Diego Marcilio and
               Rodrigo Bonif{\'a}cio and
               Eduardo Monteiro and
               Edna Canedo and
               Welder Luz and
               Gustavo Pinto},
  title     = {Are Static Analysis Violations Really Fixed? {A} Closer Look at
               Realistic Usage of {SonarQube}},
  booktitle = {Proceedings of the 27th International Conference on Program
               Comprehension},
  series    = {ICPC '19},
  pages     = {209--219},
  year      = {2019},
  publisher = {IEEE Press},
  doi       = {10.1109/ICPC.2019.00040}
}

@misc{lin2025deprecated,
  author        = {Guancheng Lin and Xiao Yu and Jacky Keung and
                   Xing Hu and Xin Xia and Alex X. Liu},
  title         = {Lightweight Model Editing for {LLMs} to Correct
                   Deprecated {API} Recommendations},
  year          = {2025},
  eprint        = {2511.21022},
  archivePrefix = {arXiv},
  primaryClass  = {cs.SE},
  url           = {https://arxiv.org/abs/2511.21022},
  note          = {arXiv preprint}
}

@misc{superblocks2025lovable,
  author       = {{Superblocks}},
  title        = {Lovable Vulnerabilities: {A} Security Analysis},
  howpublished = {\url{https://www.superblocks.com/blog/lovable-vulnerabilities}},
  year         = {2025},
  note         = {Accessed 2026-05}
}

@misc{cve2025lovable,
  author       = {{MITRE Corporation}},
  title        = {{CVE-2025-48757}},
  howpublished = {\url{https://www.cve.org/CVERecord?id=CVE-2025-48757}},
  year         = {2025},
  note         = {Accessed 2026-05}
}

@misc{sonarqube_docs,
  author       = {{SonarSource}},
  title        = {{SonarQube} Server Documentation --- Issues},
  howpublished = {\url{https://docs.sonarsource.com/sonarqube-server/latest/user-guide/issues/}},
  year         = {2025},
  note         = {Accessed 2026-05}
}

\end{document}